\documentclass[useAMS]{mn2e}
\usepackage{graphics,times,epsfig}

\usepackage{times}

\newcommand{\be}{\begin{equation}}
\newcommand{\ee}{\end{equation}}

\newcommand{\msun}{{$M_{\odot}$}}

\newcommand{\es}{erg~s$^{-1}$}

\newcommand{\esc}{erg~s$^{-1}$~cm$^{-2}$}
\newcommand{\degree}{$^{\circ}$} 
\newcommand{\gtsima}{$\; \buildrel > \over \sim \;$}
\newcommand{\ltsima}{$\; \buildrel < \over \sim \;$}
\newcommand{\prosima}{$\; \buildrel \propto \over \sim \;$}
\newcommand{\gsim}{\lower.5ex\hbox{\gtsima}}
\newcommand{\lsim}{\lower.5ex\hbox{\ltsima}}
\newcommand{\simgt}{\lower.5ex\hbox{\gtsima}}
\newcommand{\simlt}{\lower.5ex\hbox{\ltsima}}
\newcommand{\simpr}{\lower.5ex\hbox{\prosima}}

\newcommand{\etal}{{et al.~}}
\newcommand{\cxo}{\textit{Chandra}}

\newcommand{\logedd}{log($L_{\rm X}$/$L_{\rm Edd}$)}

\newcommand{\so}{V4641~Sgr}

\def\ciao{\hbox{\rm{\small CIAO}}}

\newcommand{\rev}{{Revnivtsev}}
\newcommand{\apj}{{\it ApJ}}

\title[A microblazar in V4641~Sgr]{V4641 Sgr: a candidate precessing microblazar}
\author[E. Gallo \etal]{Elena Gallo$^{1,2}$\thanks{E-mail: egallo@umich.edu}, Richard M. Plotkin$^{1}$, Peter G. Jonker$^{3,4,5}$\\\\
$^{1}${Department of Astronomy, University of Michigan, 500 Church Street, Ann Arbor, MI 48109, USA}\\
$^{2}${Kavli Institute For Theoretical Physics, Kohn Hall, University of California Santa Barbara, CA~93106, USA}\\
$^{3}${SRON, Netherlands Institute for Space Research, Sorbonnelaan 2, 3584~CA, Utrecht, NL}\\
$^{4}${Harvard--Smithsonian  Center for Astrophysics, 60 Garden Street, Cambridge, MA~02138,USA}\\
$^{5}${Department of Astrophysics/IMAPP, Radboud University Nijmegen,
P.O.~Box 9010, 6500 GL, Nijmegen, NL}}
\begin{document}
\maketitle
\begin{abstract}
The X-ray spectrum of the Galactic X-ray binary \so\ in outburst has been found to exhibit a remarkably broad emission feature above 4 keV, with inferred equivalent widths up to 2 keV. Such a feature was first detected during the X-ray flaring activity associated with the giant outburst that the source experienced in 1999 September. The extraordinarily large equivalent width line was then ascribed to reflection/reprocessing of fluorescent Fe emission within an extended optically thick outflow enshrouding the binary system as a result of a short-lived, super-Eddington accretion episode. 
Making use of new and archival X-ray observations, we show here that a similar feature persists over four orders of magnitude in luminosity, down to Eddington ratios as low as \logedd$\simeq-4.5$, where the existence of an optically thick envelope appears at odds with any viable accretion flow model. 
Possible interpretations for this highly unusual X-ray spectrum include a blend of Doppler shifted/boosted Fe lines from unresolved X-ray jets ($a~la$ SS433), or, the first Galactic analog of a blazar spectrum, where the $>$ 4 keV emission would correspond to the onset of the Inverse Compton hump. Either requires a low inclination angle of the jet with respect to the line of sight, in agreement with the estimates for the 1999 superluminal jet~($i_{\rm jet}<10$\degree). 
The fast variability of the feature, combined with the high orbital axis inclination (60\degree$<i_{\rm orb}<$71\degree), argue for a rapidly precessing accretion flow around \so, possibly leading to a transient {\it microblazar} behavior. 
\end{abstract}


\section{Introduction}
\label{sec:intro}

The Galactic transient \so\ is perhaps best known for the bright 1999 outburst that led to the identification of its X-ray counterpart (SAX J1819.3-2525; Markwardt \etal 1999; in'~t~Zand \etal 1999), and for being one of just a handful of Galactic X-ray binary systems to have shown apparent superluminal radio ejecta (Hjellming \etal 2000). It has since been established to host a strong black hole candidate ($9.6^{+2.1}_{-0.9}$ \msun) accreting from a high mass ($6.5^{+1.6}_{-1.0}$ \msun) companion star at a most likely distance of 9.6 kpc (Orosz \etal 2001). 
The 1999 outburst was characterized by particularly intense activity at all wavelengths over September 14-15. Three X-ray flares were detected during the rising and decaying parts of the optical light curve, with peak X-ray luminosities reaching or exceeding the Eddington limit for the system (Revnivtsev \etal 2002a). The source X-ray spectrum, as measured by the {\it Rossi} {X-ray Timing Explorer} during one of the intra-flare minima, exhibited a remarkable emission feature ascribed to Fe line emission centered at 6.63$\pm$0.08 keV, and with an equivalent width of 2.4$\pm$0.3 keV. The exceptional broadening of the line was ascribed to reprocessing/reflection within an extended optically thick envelope/outflow enshrouding the system as a result of a brief super-Eddington accretion episode (Revnivtsev \etal 2002a, 2002b). The presence of a high velocity wind associated with the September 1999 X-ray flaring activity was also suggested based on optical and near-infrared observations (Charles \etal 1999; Chaty \etal 2003).
A similar behavior -- including a broad emission line in the fluorescent Fe energy range -- was described by Maitra \& Bailyn (2006) reporting on coordinated optical and X-ray observations of the 2003 outburst of \so. In that case, the inferred Fe line equivalent width was close to 1 keV.   
In this Letter, we present and discuss the enigmatic spectrum of \so\ as measured by the \cxo\ X-ray Observatory during three visits over 2004 July and August, as part of a Target of Opportunity campaign to follow the return of a black hole X-ray binary system into quiescence. The observations were triggered following a brief period of increased activity reported in early July 2004 by Swank \etal (2004). As detailed in Section 2, the source X-ray luminosity during the first visit was already no higher than a few $10^{34}$ \es, confirming that \so\ was indeed fading towards quiescence. Despite the low luminosity level, the \cxo\ spectrum still exhibits a remarkably broad emission feature above 4 keV. 
Notwithstanding the severe limitations inherent to lack of data above 10 keV, the nature of this feature is discussed in the context of previous observations of the same source, as well as in the more general perspective of X-ray observations of hard and quiescent state black hole X-ray binaries. 
\begin{figure*}
\epsfig{figure=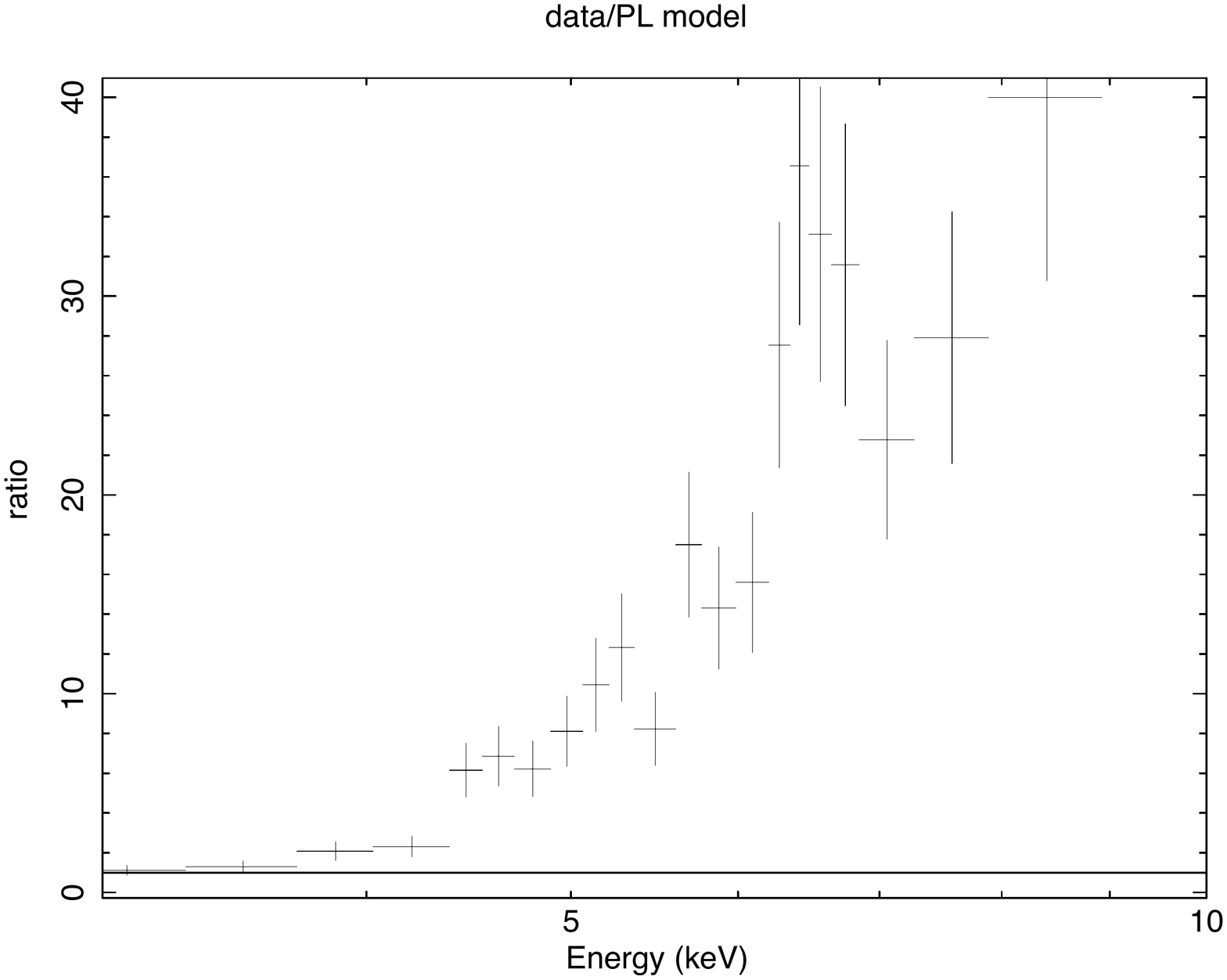,width=0.25\textwidth,angle=-90}\hspace{-0.2cm}
\epsfig{figure=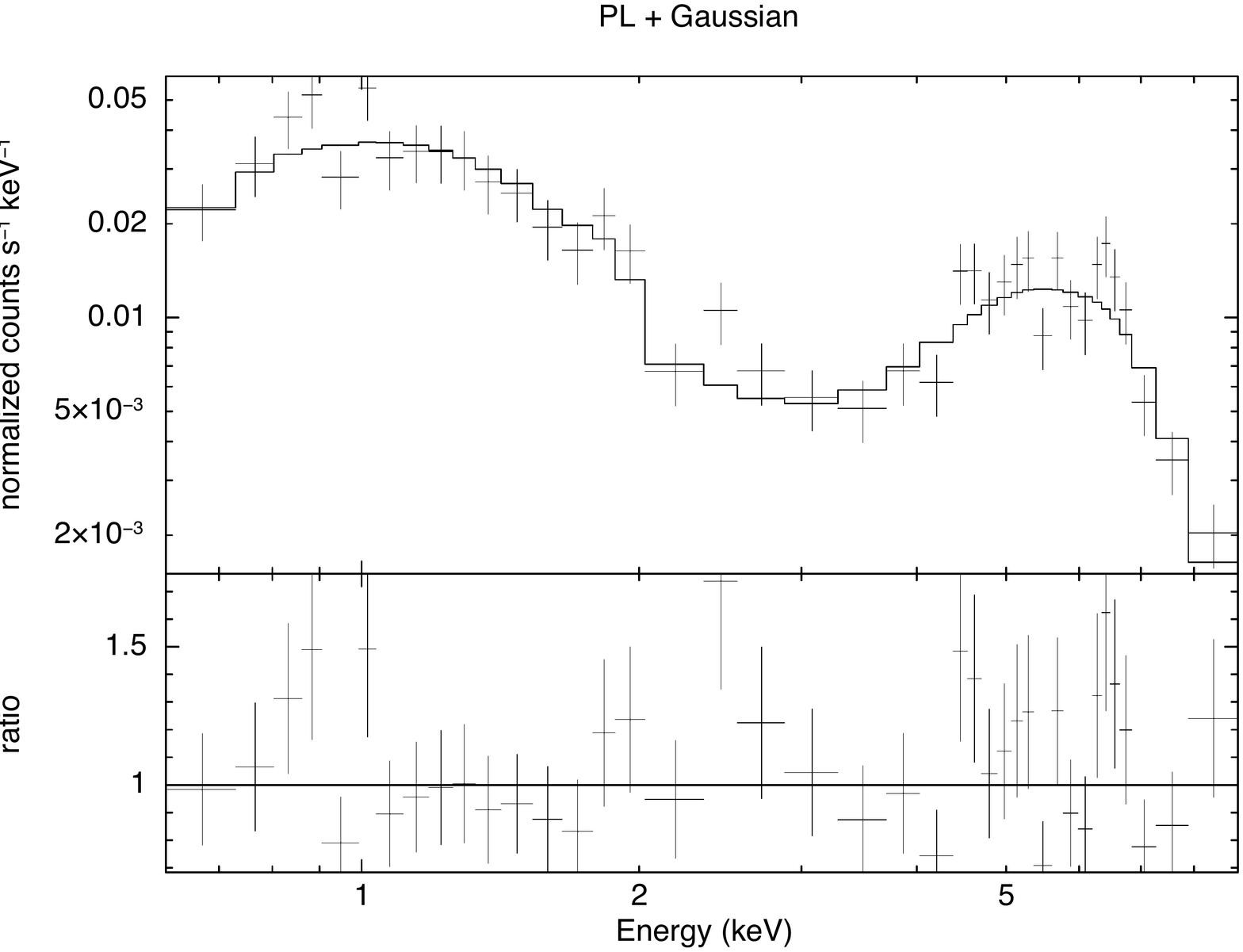,width=0.25\textwidth,angle=-90}
\epsfig{figure=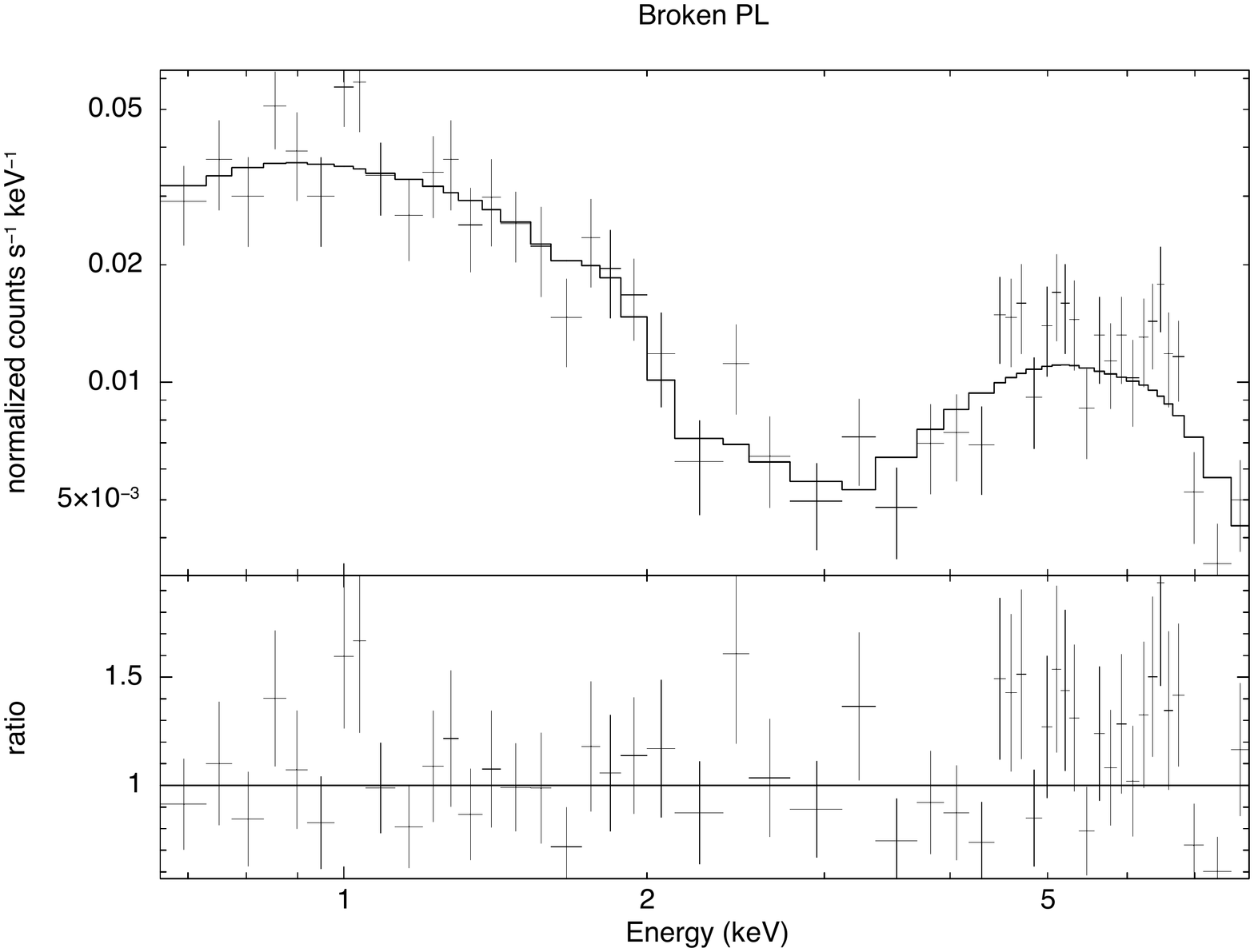,width=0.25\textwidth,angle=-90}
\caption{\cxo\ ACIS-S spectrum of \so\ as observed on 2004 July 17 (ObsID 4451). Left: Data/model ratio for an absorbed power law model with the 4-7 keV energy range was ignored in the fitting. A prominent emission component is apparent above 4 keV, possibly indicative of reflection and Fe K line emission. Middle \cxo\ ACIS-S spectrum of \so\ as measured on 2004 July 17, and fitted with an absorbed power law model plus a broad Gaussian profile. The bottom panel shows the data/model ratio for the best-fitting model shown above. Right: Same as middle, where the spectrum is fitted with an absorbed broken power law model.\label{fig:best} }
\end{figure*} 
\section{Data analysis}
\label{sec:data}

\so\ was observed by \cxo\ ACIS-S on 2004 July 17, for approximately 10 ks (ObsId 4451). This was the first of three triggered observations (PI Jonker), to be followed by ObsId 4452 (20 ks) and ObsId 4453 (40 ksec), on 2004 July 30 and August 10, respectively. 
New evt2 files were created and calibrated using the \ciao\ (version 4.5) script {\rm {\small chandra$\_$repro}}, which also cleans the ACIS particle background for very faint mode and generates a new bad pixel file. The observation light-curves were inspected for background flares (none were found), yielding net exposure times of 9.2, 18.3 and 36.4 ksec for ObsId 4451, 4452 and 4453, respectively. Further analysis was restricted to energies between 0.3 and 10 keV. 
An X-ray source is detected at a position consistent with \so\ in all three observations. Net count rates were estimated using a circular source extraction region centered at the source nominal position and with a radius of 10\arcsec, while an annulus with inner and outer radius of 25\arcsec\ and 35\arcsec\ was chosen for the background extraction. The measured average net count rates were $9.59\pm 0.32 \times 10^{-2}$, $1.37\pm 0.28 \times 10^{-3}$ and $9.0\pm 1.6\times 10^{-4}$ cps, confirming the steep decline of \so\ into quiescence. 
With a total of 877 net counts on source, ObsId 4451 (2004 July 17) is the only one to allow for a proper spectral fitting. 
We first focus on this observation.
The source and background spectrum were extracted from the same regions as mentioned above making use of the \ciao\ script {\rm{\small specextract}}. The resulting energy spectrum, grouped to have a minimum of 20 counts per bin, was fitted within XSPEC version 11.3 (unless otherwise noted, all the errors are quoted at the 90 per cent confidence level). 
In line with previous works reporting on non flaring X-ray observations of \so, we fixed the equivalent neutral hydrogen column density to 0.23$\times 10^{22}$ atoms cm$^{-2}$, as given by Dickey \& Lockman (1990). As is customary, we first attempted to fit the continuum spectrum with (i) a simple power law model; (ii) a multi-colour disc blackbody ({\rm{\small diskbb}}) and a two-parameter Comptonization model ({\rm{\small simpl}}) convolved with a multi-colour disc, all attenuated by intervening absorption ({\rm{\small phabs}}). None of the above models gave a statistically acceptable fit, with $\chi^2 / \nu = 284/38, 287/39$, and $335/35$, respectively  (where $\nu$ denotes the number of degrees of freedom). Similar results were obtained by allowing the equivalent neutral hydrogen column to vary (in which case, the best fit values were consistent with the Galactic value). 
The ratio of the spectrum to a power law model with slope $\Gamma=2$ (typical of quiescent black hole X-ray binaries; Plotkin \etal 2013) is shown in the left panel of Figure 1, to highlight the presence of a remarkably broad feature above 4 keV, possibly indicative of Fe K line emission and/or reflection.
To investigate the nature of this broad emission component, we first adopted a model consisting of an absorbed power law plus an emission line with a Gaussian profile (Figure 1, middle). With this addition, the fit improves substantially, to $\chi^2 / \nu = 38/35$. The best fit centroid energy is measured to be $E=7.15^{+0.74}_{-0.44}$ keV (consistent with highly ionized Fe), the inferred line width is $\sigma = 1.73^{+0.44}_{-0.29}$ keV, while the underlying power law continuum has an index $\Gamma=2.32\pm0.28$.  The measured unabsorbed flux in the 0.3--10.0 keV range is $4.3\times 10^{-12}$ \esc. Adopting the distance and black hole mass estimated by Orosz \etal (2001), this corresponds to a luminosity $L_{\rm X}=4.8\times 10^{34}$ \es, or \logedd~$\simeq -4.5$ for a 9.6 \msun\ black hole. 
Visual inspection of the data/model ratio (bottom panel of the middle plot in Figure 1) might suggest the presence of a narrow emission feature at $\sim 6.4$ keV. Indeed, fitting the spectrum with an absorbed power law model plus two Gaussian profiles returns $\chi^2 / \nu = 31/32$. This is achieved by having a broad ($\sigma = 1.90^{+0.69}_{-0.37}$ keV) line centered at $E=7.30^{+1.20}_{-0.58}$ keV plus a narrower component ($\sigma = 0.14\pm0.14$ keV) at $E=6.47\pm0.14$ keV. From a statistical point of view however, the addition of the second Gaussian is not required by the data. 
We attempted to substitute the broad Gaussian component with a  {\rm\small{laor}} profile, allowing the line energy to vary between 6.40 and 6.97 keV (spanning the energy range from neutral to hydrogenic Fe).  While a statistically acceptable fit is found ($\chi^2 / \nu = 36/33$) for a line energy $E=6.44^{+0.45}_{-0.04}$ keV,  the line emitting region is completely unconstrained, with the inner and outer radius covering the full range allowed by the model, i.e. between 1 and 400 gravitational radii. The relatively low number of counts and the lack of data above 10 keV prevent us from exploring sophisticated model with a large reflection component.
Lastly, the shape of the $>4$ keV residuals in the left plot of Figure 1 could be well approximated simply by a second (rising) power law component; however, fitting the overall spectrum with an absorbed broken power law model ({\rm{\small bknpower}}; Figure 1, right plot) yields a worse results ($\chi^2/ \nu=50/35$) compared to the broad Gaussian plus single power law model (middle panel of Figure 1). 

\begin{figure}
\epsfig{figure=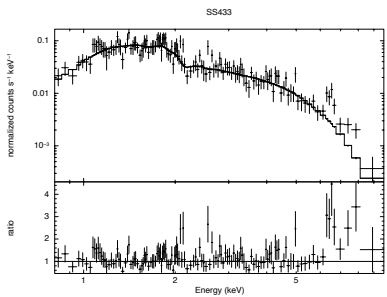,width=0.25\textwidth,angle=-90}
\caption{Top: Cumulative core+jet spectrum of the Galactic jet source SS~433 (ObsId 659), fit by an absorbed power law model fit below 5 keV. Bottom: While the red- and blue-shifted lines are clearly visible in the data/model ratio, their contribution to the spectrum is not nearly as significant as the excess emission seen in \so\ (left panel of Figure 1).
 \label{fig:ss433}}
\end{figure} 
Given the unusually broad nature of this emission component, it is worth investigating whether it might be related to poor background subtraction or have a different origin than the black hole X-ray binary system itself (e.g. the Galactic ridge emission is known to have a prominent emission line at 6.7 keV; \rev\ \etal 2007). To do so, we followed the approach of Tomsick \etal (2009), who tested for the presence of a narrow 6.7 keV line in the 2008 September X-ray spectrum of the black hole X-ray binary GX~339--4. We first examined the spectrum of the ACIS background region, and we found no evidence for an emission line. We then extracted a new spectrum of \so\ within a circular region of 2\arcsec\ in radius (vs. 10\arcsec\ in the previous extraction), thereby reducing the background by a factor 25. Fitting the new source spectrum with an absorbed power law model still shows prominent residuals above 4 keV. When a Gaussian is added to the model, the fitted centroid energy is $E=6.99^{+0.79}_{-0.41}$ keV, and the model is consistent with the same power law to Gaussian normalization ratio as measured for the larger extraction region. We are thus drawn to the conclusion that the broad emission feature is intrinsic to the spectrum of \so.

Next, we report on \cxo\ observations at different luminosity regimes, and we show that \so\ is characterized by an unusually hard X-ray spectrum down to the lowest luminosity level.  
We extracted spectra from ObsIDs 4452 and 4453 over 2$\arcsec$ circular apertures and fit an absorbed power-law to each using Cash statistics.  Again fixing the column density to $0.23 \times 10^{22}$ cm$^2$, the best-fit slopes are $\Gamma = 0.93\pm0.27$ and $1.07\pm0.25$  (for reference, a single power law fit to the spectrum of ObsID 4451 yielded $\Gamma=0.92\pm0.09$). All three observations -- spanning Eddington ratios between $-7<$\logedd$< -4.5$ -- are much harder than typical quiescent black hole X-ray binaries (e.g. Corbel \etal 2006; Plotkin et al. 2013, and references therein), suggesting that the broad emission feature reported above persists all the way through quiescence.
This is confirmed by archival \cxo\ observations taken in 2002 (ObsIds 3800 and 4285), and discussed by Tomsick et al. (2003). The combined observation, totaling  29.5 ksec of net exposure, yielded a source with 9 counts at a position consistent with the radio counterpart, implying a net count rate comparable to the lowest reported above (ObsId 4453). The photon index measured by Tomsick \etal (also using Cash statistics and an absorbed power law model) is $\Gamma= 0.2^{+0.9}_{-0.1}$ for V4641 Sgr, in line with the hard values reported in this work for the 2004 observations. 

\section{Interpretation and comparison with SS433}

\so\ appears to be a unique system, in that its X-ray spectrum {\it in quiescence} (cf. McClintock \& Remillard 2003) exhibits a prominent emission feature above 4 keV.  Whatever its origin, this feature seems to be persistent across at least four orders of magnitude in luminosity, and possibly accretion rate. 
While coverage above 10 keV is obviously necessary in order to properly model the continuum, there is no obvious explanation for the observed emission feature. Neglecting relativistic effects, line widths in the measured range can only be achieved for scattering/reprocessing within a very optically thick medium (optical depth of the order $\sim 3$; Kallman \& White 1989). While this may well be the case for brief episodes of super-Eddington accretion (\rev\ \etal 2002a), such high optically depths are difficult to reconcile with Eddington ratios as low as $-4.5$ (this work; Tomsick \etal 2003). Additionally, even if a dense cloud were still present, it would only be partially ionized at such low luminosities, making Compton scattering very inefficient at smearing out the line. 

One captivating possibility is that the broad feature arises in a X-ray emitting jet, similar to that inferred from low spatial resolution observations of SS443 (Margon 1984). In fact, the similarity of the 1999 September spectrum of \so\ with SS~433's was already pointed out by Mart\'i \etal (2001) and Chaty \etal (2003), both reporting on the 1999 activity. 
In order to test whether the X-ray emission from \so\ could be due to the superposition of broad, Doppler shifted Fe emission lines, of the same kind as those detected in the (spatially resolved) X-ray jets of SS433 (Migliari \etal 2002, 2005), we extracted the spectrum of SS433 as observed by \cxo\ ACIS-S in 2001 (ObsId 659). Unlike Migliari et al. (2002; see figure 2 for comparison and for a description of the kinematic model), we extracted the spectrum from the entire X-ray source associated with SS433, including the core and the two resolved jets\footnote{Though the core of SS433 suffered from heavy pile-up in these observations, this does not affect the results of our qualitative comparison.}. The spectral extraction was carried out following the same steps described above for \so, except an ellipse with a 15 arcsec semi-major axis aligned with the resolved X-ray jets orientation axis was adopted for the source; the semi-minor axis was 10 arcsec. The 0.3-10 keV spectrum of SS433 is shown in Figure~\ref{fig:ss433}, to be compared with \so's (caveat more than three orders of magnitude difference in Eddington scaled luminosities). While it still shows a prominent excess (with respect to a power law fit) in the 6-8 keV range -- known to be due to the combination of blue- and red-shifted Fe emission lines from the jets -- the continuum component over-shines the lines and makes their emission much less prominent if compared to the spectra of the East and West jets discussed by Migliari et al. (2002). 
This qualitative comparison illustrates that, if the unusually hard X-ray spectrum of \so\ is indeed due to emission from X-ray jets, their contributions relative to the core has to be substantially higher compared to the case of SS433, possibly due to stronger Doppler boosting effects. 
This raises the interesting possibility that \so\ might be a Galactic {\it microblazar}, i.e. a stellar mass analog of blazar sources, whose relativistic jets are very closely aligned with our line of sight. Interestingly, the microblazar scenario was already proposed for \so\ by Orosz \etal (2001) and further elaborated by Chaty \etal (2003) on the basis of the source optical and near-infrared properties during the giant 1999 outburst. In particular, Chaty \etal (2003) note that the lack of Doppler-shifted H$_{\alpha}$ lines in their optical spectra would still be consistent with a highly beamed emission. In fact, assuming a jet intrinsic velocity of 0.95$c$ and $i_{\rm jet}<10$\degree~(as inferred for the 1999 superluminal radio jet), the approaching and receding line components would have been blue- and red-shifted to UV ($\sim$ 1000 \AA) and near-infrared ($\sim$4 $\mu$m) wavelengths, respectively, and thus missed. 
\begin{figure}
\vspace{-0.5cm}
\epsfig{figure=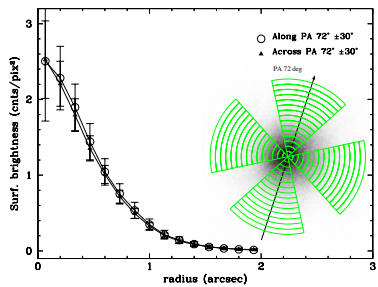,width=0.35\textwidth,angle=0}
\vspace{-0.25cm}
\caption{Radial surface brightness profile of the sub-pixelated image of \so\ along (open circles) and across (filled triangles) the position angle of the extended radio source associated with its 1999 outburst (Hjellming \etal 2000). No extended X-ray emission is detected in either direction.  \label{fig:res}}
\end{figure} 
Since, however, optical and infrared emission are thought to be produced downstream the jet, where the emitting particles are likely to have suffered significant deceleration compared to the X-ray emitting particles, this scenario would imply an higher yet intrinsic velocity (and thus narrower beaming cone) for the X-ray emitting nozzle at the base of the jet. In turn, this requires a scenario whereby the observer's line of sight falls outside the (narrower) X-ray beaming cone and yet inside the (wider) optical/infrared cone (we note, however, that it is the combination of inclination angle and Lorentz factor that determines whether the emission is boosted or de-boosted, and it would still be possible to accommodate various scenarios with different permutations of those two parameters).
A very low inclination angle with respect to our line of sight is not in contrast with estimates based on the apparent superluminal ejecta detected from \so\ during its 1999 outburst ($i_{\rm jet}<10$\degree; Orosz \etal 2001). This is, however, in contrast with the inferred inclination of the binary orbital inclination, for which Orosz \etal estimate $60$\degree$< i_{\rm orb}< 70$\degree, raising the interesting possibility of a {\it precessing} inner accretion disc/jet (see Martin, Reis \& Pringle 2008 for a discussion on the relative alignment of the inner/outer accretion disc and the jet). 

In order to achieve a higher spatial resolution and search for possible extended X-ray emission from \so, we first reprocessed the event file (ObsId 4451) with the energy-dependent sub-pixel event-repositioning (EDSER) algorithm (Li \etal 2004) using the {\rm{\small chandra$_{-}$repro}} \ciao\ script. 
We simulated the best available point spread function (PSF) for this specific observation using the Chandra Ray Tracer (ChaRT) package, where the input spectrum was extracted from ObsId 4451 as described above. A sub-pixel image of the PSF was then created from the ChaRT-generated (pseudo) event file using the MARX software (version 5.0) to project the rays onto the detector plane.
Having no pixel quantization, the projected event file created with ChaRT and MARX can binned by an arbitrary value to create a sub-pixelated image, which can then be convolved with the raw image. The resulting sub-pixelated image of \so, with 1/10 of the native pixel resolution, is shown in the inset of Figure~\ref{fig:res}, and does not display any obvious extension. To quantify this, we extracted the surface brightness profiles along and across the inferred P.A. of the radio jet of \so\ (Hjellming \etal 2000; Orosz \etal 2001), adopting slices with 30\degree\ semi aperture, as shown in Figure~\ref{fig:res}; this analysis confirms the lack of any detectable excess X-ray emission along the radio jet axis. While we would not necessarily expect to detect extended X-ray emission from the jets if they were indeed aligned within 10\degree~of our line of sight, whether extended X-ray emission could be detected depends on many factors other than inclination, such as cooling time, jet speed and opening angle, all of which are relatively uncertain.

To summarize, the X-ray spectrum of \so\ exhibits a remarkably broad emission feature centered at around 6.5 keV (albeit with large variations in centroid energy and equivalent width across different epochs). Earlier work, often based on higher quality spectra -- in that they were taken with instruments with wider energy coverage, such as {\textit{Beppo}}-SAX or RXTE, as well as at $\simgt 10-100$ times higher flux levels --- ascribed the broad feature to reprocessing/reflection of fluorescent Fe emission within an optically thick outflow triggered by an episode of super-Eddington accretion (e.g. \rev\ \etal 2002a, 2002b for the 1999 outburst; Maitra \& Bailyn 2006 for the 2003 outburst, dwells 1,3 and 4). Corroborating evidence for a high velocity outflow during the 1999 outburst came from optical analysis as well (Charles \etal 1999; Chaty \etal 2003).
Although it appears to be persistent over a wide range in Eddington ratios, RXTE observations carried out during the 1999 September outburst also indicate that the nature of the broad emission feature is variable, possibly due to the interaction with high velocity ejecta/wind (see e.g. Chaty \etal 2003 and Maitra \& Bailyn 2006). In addition, while in'~t Zand (2000) reported on the detection of a broad Fe line in the (pre-giant flare)  February 1999 {\it Beppo}-SAX spectrum of \so, a subsequent re-analysis ascribed the broadening to relativistic effects (Miller \etal 2002) . This would indicate that there might be multiple processes at work in shaping the $>4$ keV emission spectrum of \so. 
It is interesting to note, however, that the best-fit reflection model for the the February 1999 {\it Beppo}-SAX observation implies an {\it inner} disc inclination ($i_{\rm disc}=43$\degree$\pm 15$\degree; Miller \etal 2002) that is not consistent with the inferred radio jet inclination ($i_{\rm jet} <10$\degree; Hjellming \etal 2000; Orosz \etal 2001), and is only marginally consistent with the orbital inclination (60\degree$<i_{\rm orb}<$71\degree). 

This complex pattern of behavior points toward a highly dynamic scenario in which the broad emission feature -- whatever its origin -- can vary on relatively short (days, possibly shorter) time-scales, although it persists from Eddington-limited luminosities down to quiescence.
The hypothesis that the broad line detected in \so\ at \logedd$\simeq -4.5$ (this work) may be due to {\it relativistically} broaden Fe emission arguably raises more issues than it actually solves, as the survival of a Keplerian disc that extends down to the innermost stable circular orbit at these luminosity levels appears unphysical. Similarly, reprocessing within an optically thick wind, as suggested for the 1999 September outburst, seems highly unlikely. 
Hence, the survival of the broad feature in quiescence requires a novel interpretation. The qualitative similarity with the SS4433 X-ray spectrum suggests that \so\ might be a more extreme (and transient) version of the well known Galactic jet source. For instance, if \so\ hosted {\it precessing jets} that were closely aligned to our line of sight only over a fraction of their precession period, the source could appear as a variable microblazar, and its broad $>4$ keV emission be due to a blend of highly Doppler boosted Fe lines from the X-ray emitting jets. In this scenario, not only is rapid time-scale variability of the broad feature itself expected, but could probe the typical precession jet period. Rapid variability is indeed detected even in our 2004 July 17 observation; the 0.3-10 keV light curve exhibits a bright (factor $\sim 10$) flare during the first 2 ksec. While the low number of counts prevents us from extracting a meaningful spectrum for the post-flare segment of the observation, we notice that the [4.0-10 keV]/[0.5-4.0 keV] count ratio during the flare is somewhat harder than the post-flare ratio (i.e. $1.03\pm0.25$ and $0.46\pm0.19$, respectively). 
If this short time-scale variability were indeed related to the system precession period, this would imply a fairly compact jet launching region. 
The Lense-Thirring precession period for a compact binary system can be parametrized as a function of the compact object mass, spin parameter $a_{\star}$ and radial distance from the rotating compact object. Following equation 4 of Massi \& Zimmermann (2010), a 2,000 sec precession period would translate into a radial distance between ~110(235) gravitational radii for a spin parameter $a_{\star}$=0.1(1).  
We note, however, that transient relativistic jets, typically associated with X-ray outbursts, are generally thought (albeit not conclusively proven) to be more relativistic than persistent jets associated with the hard and quiescent X-ray states (e.g. Gallo \etal 2003, Fender Belloni \& Gallo 2004). If this were the case for \so, then we would expect the EW of the line to decrease with the overall bolometric luminosity, in contrast with the data. 

An alternative interpretation, mainly inspired by the fact that the data/model ratio presented in the left panel of Figure 1 appears to be a smooth power law above 4 keV, is that the X-ray spectrum of \so\ might represent the first Galactic analog of a blazar spectrum (e.g. Fossati \etal 1998), in which the broad emission feature constitutes the rising part of the so called Inverse Compton hump (as for the Fe line blend interpretation, the hump would only emerge when the jet is closely aligned to the line of sight).  
We explored this scenario in more detail following Ghisellini \& Tavecchio (2009). For canonical blazar sources, the Compton hump arises either from Synchrotron-Self-Compton (SSC) or External Compton (EC) scattering.  In the former case, the jet synchrotron radiation responsible for the low energy spectral peak (typically observed in the sub-mm/IR bands) is up-scattered by the same population of relativistic electrons that produces the seed synchrotron radiation in the first place. In the latter, the seeds photons for the inverse-Compton scattering process are provided by an external source.  
The characteristic peak energy/frequency of the Compton hump is simply set by the characteristic peak energy of the seed photons and the Lorentz factor of the electron population, $\gamma_e$. Specifically, assuming -- by analogy with other black hole X-ray binaries -- that the jet synchrotron emission peaks in the IR band (i.e. $\simeq 2\times 10^{13}$ Hz), the SSC scenario requires fairly highly relativistic electrons, with $\gamma_e = ({2\times 10^{19} \rm Hz}/{10^{13}\rm Hz })^{1/2}\simeq 1000$.  
While this is not necessarily unreasonable, we have no direct observational constraints on the spectral extent of the radio counterpart to \so, as its sub-mm to IR spectrum is dominated by the donor star.    
Instead, in the case of EC, the massive companion star (spectral type B9) could provide the external photon field. By the same scaling arguments as above, since the stellar emission peaks in the UV band (at $\simeq 5\times 10^{15}$ Hz), EC would require only mildly relativistic Lorentz factors, i.e. $\gamma_e \simeq 1-100$. However, for EC to dominate over SSC, the (co-moving) energy density in external radiation ($u'_{\rm r}$) must exceed the energy density in magnetic field ($u'_{\rm B}$). Thus, the energy density in external radiation can be written as $u_{\rm r}=L_{\star}/(4\pi R^2_{\rm orb}c)$, where $L_{\star}$ is the donor star luminosity, $R_{\rm orb}$ is the system orbital separation and $c$ is the speed of light. For the magnetic energy density, assuming that the jet Poynting flux is constant and its value is well approximated by the system bolometric luminosity ($L_{\rm bol}$), we have that  $u'_{\rm B} = L_{\rm bol}/(\pi R^2_{\rm diss}\Gamma^2c)$, where $\Gamma$ is the jet bulk Lorentz factor and $R_{\rm diss}$ is the distance between the jet base and the location of the magnetic energy dissipation region. Hence, substituting $\Gamma^2u_{\rm r}$ for $u'_{\rm r}$, the requirement that $u'_{\rm r}>u'_{\rm B}$ translates into: $(L_{\rm bol}/L_{\star})<(\Gamma^4/4)(R_{\rm diss}/R_{\rm orb})^2$. Plugging in the appropriate values for $L_{\star}$ and $R_{\rm orb}$ as measured for \so\ by Orosz \etal (2001), and assuming $L_{\rm bol}\simeq L_{\rm X}=4.8\times 10^{34}$ \es, this returns a minimum bulk Lorentz factor between 15-45 for a dissipation radius between 100-1000 Schwarzschild radii. However, we wish to stress that the minimum Lorentz factor increases with $L_{\rm bol}$. In addition, bulk Lorentz factors as high as 10-50 seem to be more typical of super-massive black holes in blazars than X-ray binary systems (e.g. Fender \etal 2004). 
While not unphysical, the Compton hump interpretation is admittedly highly speculative, particularly as it stems from observations based on such a narrow spectral energy range. Future high sensitivity hard X-ray observations of this intriguing source, e.g. by NuSTAR, will possibly shed light on the nature of the bizarre spectrum of \so. 
\section*{Acknowledgments}
This research was supported in part by the National Science Foundation under Grant No. NSF PHY11-25915. 
EG would like to thank Mike Eracleous, Gabriele Ghisellini and Francesco Haardt for helpful suggestions.

\end{document}